\documentclass[letterpaper,11pt]{article}
\usepackage{authblk}
\usepackage[margin = 1in]{geometry}
\usepackage{graphicx}
\graphicspath{{Figures/}{logo/}}
\usepackage{amsmath,amssymb}
\usepackage{mathtools}
\usepackage{algorithm, algorithmic, float}
\usepackage{tikz}
\usetikzlibrary{positioning,calc}
\usepackage{amsthm}
\usepackage{array, multirow, booktabs}

\newcommand{\ket}[1]{|#1\rangle}
\newcommand{\bra}[1]{\langle#1|}
\newcommand{\x}{\otimes}
\newcommand{\Reg}{\mathsf{R}}
\newcommand{\Heg}{\mathrm{HT}}
\newcommand{\spn}{\mathrm{span}}
\newcommand{\Step}[1]{step.~\hyperref[step:#1]{\ref*{step:#1}}}
\newcommand{\tab}[1]{Table~\hyperref[tab:#1]{\ref*{tab:#1}}}

\newtheorem{remark}{Remark}
\newtheorem{theorem}{Theorem}
\newtheorem{lemma}{Lemma}

\usepackage[colorlinks = true]{hyperref}
\usepackage{cleveref}
\title{%Improved 
	Faster quantum mixing of Markov chains in non-regular graph with fewer qubits}
\author[a,b]{Xinying Li}
\author[a,c]{Yun Shang \thanks{shangyun@amss.ac.cn}}
\affil[a]{Institute of Mathematics, Academy of Mathematics and Systems Science, Chinese Academy of Sciences, Beijing 100190, China}
\affil[b]{School of Mathematical Sciences, University of Chinese Academy of Sciences, Beijing 100049, China}
\affil[c]{NCMIS, MDIS, Academy of Mathematics and Systems Science, Chinese Academy of Sciences, Beijing 100190, China}

\date{}
\begin{document}
	\maketitle
	\begin{abstract}				
		Sampling from the stationary distribution is one of the fundamental tasks of Markov chain-based algorithms and has important applications in machine learning, combinatorial optimization and network science. 
		For the quantum case, qsampling from Markov chains can be constructed as preparing quantum states with amplitudes arbitrarily close to the square root of a stationary distribution instead of classical sampling from a stationary distribution.
		In this paper, a new qsampling algorithm for all reversible Markov chains is constructed by discrete-time quantum walks and works without any limit compared with existing results.
		
		In detail, we build a qsampling algorithm that not only accelerates non-regular graphs but also keeps the speed-up of existing quantum algorithms for regular graphs. In non-regular graphs, the invocation of the quantum fast-forward algorithm accelerates existing state-of-the-art qsampling algorithms for both discrete-time and continuous-time cases, especially on sparse graphs. Compared to existing algorithms we reduce $\log n$, where $n$ is the number of graph vertices.
		In regular graphs, our result matches other quantum algorithms, and our reliance on the gap of Markov chains achieves quadratic speedup compared with classical cases. 
		For both cases, we  reduce the number of ancilla qubits required compared to the existing results. In some widely used graphs and a series of sparse graphs where stationary distributions are difficult to reach quickly, our algorithm is the first algorithm to achieve complete quadratic acceleration (without log factor) over the classical case without any limit.	
		
		To enlarge success probability amplitude amplification is introduced. We construct a new reflection on stationary state with fewer ancilla qubits and think it may have independent application.
		%%%%%%%%%%%%%%%%%%%%%%%%%%%%%%%%%%%%%%%%%%%%%%%%%%
		%可以作为众多机器学习或其他算法的预备处理子例程
		%%%%%%%%%%%%%%%%%%%%%%%%%%%%%%%%%%%%%%%%%%%%%%%
	\end{abstract}
	\section{Introduction}\label{intro}	
	\hspace{1.2em} 
	Random walk or Markov chains which can simulate the dynamics of a particle moving randomly on some graphs is a powerful algorithmic model for classical computer science. It can be used to solve problems in machine learning, combinatorial optimization and network science. Similarly, quantum walks which simulate the quantum coherent dynamics of a particle moving on a graph present a powerful and universal framework for designing and creating new quantum algorithms \cite{Childs2003ExponentialAS}. 
	
	Sampling from the stationary distribution of Markov chains is one of the fundamental tasks of Markov chain-based algorithms, since many problems can be reduced to sampling \cite{Aharonov2003AdiabaticQS}, such as calculating the partition function in counting problems, and estimating the importance of Internet pages (the goal of PageRank algorithm \cite{Page1999ThePC}).
	
	The classical mixing time of reversible Markov chain is bounded by $\delta^{-1}\log(\pi_{min}^{-1})$ \cite{https://doi.org/10.1112/jlms/s2-25.3.564, Jerrum2003CountingSA}, where $\delta$ denotes the eigenvalue gap of transition matrix $P$, $\pi_g$ denotes the probability of vertex $g$ in stationary distribution, $\pi_{min} \coloneqq \min_{g \in V}\pi_g$. 
	Similarly, the quantum mixing time is the minimum time for quantum walks to achieve a stationary distribution from any initial distribution, which can generally be described with an upper bound $\delta^{-1/2}\pi_g^{-1/2}$ \cite{doi:10.1137/090745854}. Although the quantum gap dependence reaches a quadratic speedup relative to the classical case, the quantum mixing time is extremely dependent on $\pi_{min}$. 
	Then quantum result largely depends on the graph size and brings extra $\sqrt{n}$ cost compared with classical case in graph with large size $n$.
	Currently, the quantum quadratic speedup of mixing time is only known for some special graphs \cite{Richter2007QuantumSO} or special stationary distributions \cite{Dunjko2015}. No universal results are obtained for more general graphs corresponding to any reversible Markov chain.
	
	Several existing results of preparing quantum stationary state in special cases are listed as follows.
	A slowly evolving Markov chain sequence is constructed in simulated annealing algorithm \cite{Wocjan2008SpeedupVQ, Somma2008QuantumSO, doi:10.1137/1.9781611975994.12}, and the total complexity is the product of $\sqrt{\delta}$ and sequence length.
 Orsucci et al. introduces a quadratic speedup algorithm that improves the additional dependency of $\pi_{min}^{-1/2}$ from $\sqrt{n}$ to $\sqrt[4]{n}$  \cite{Orsucci2018fasterquantummixing}	for those special Markov chains with slowly evolving sequence forms.
A qsampling	algorithm is prepared by reversing the search algorithm \cite{Krovi2015QuantumWC}. The qsampling time is $\Theta(\sqrt{\Heg}\frac{1}{\varepsilon})$ for any reversible Markov chain, where $\Heg$ is the classical hitting time bound. However, it depends heavily on $\varepsilon$.
	The time from single-vertex initial state $\ket{g}$ to stationary state with known $\pi_g$ is $\Theta(\sqrt{\Heg}\log\frac{1}{\varepsilon})$ in continuous-time quantum walk-based algorithm \cite{PhysRevA.102.022423}. However, since $\ket{\pi}$ is the state we want and $\pi_g$ is a component of $\ket{\pi}$, the process of preparing the value of $\pi_g$ should not be considered trivially.
%	The algorithm for reaching a long-time average probability distribution of quantum walks \cite{Chakraborty2020HowFD} prepares a limiting distribution that differs from the quantum stationary state.
	
	These results mainly focused on special Markov chains or special stationary distribution. Generally, whether the quantum stationary state can be obtained by the quadratic acceleration is still an open question.
	
	In this paper, we provide a new  discrete-time quantum walk-based qsampling algorithm in cost $\Theta(\sqrt{\Heg}\log\frac{1}{\varepsilon})$, which is suitable for any reversible Markov chain with no additional requirement on graphs or stationary distributions.
	
	Our work relies on two key technical methods. The first is the quantum interpolated walks method \cite{Krovi2015QuantumWC}, which shows how to prepare a state having constant overlap with the stationary state.
	Specifically, an appropriate interpolation parameter $s$ is chosen such that both the initial state and the stationary state have constant overlap with the 1-eigenvector (the eigenvector corresponds to the eigenvalue 1) of quantum interpolated walk operator. 
	Phase estimate can distinguish $1$-eigenvector from other eigenvectors naturally by marking ancilla registers. Combing quantum interpolated walks and phase estimate,  $\Theta(\sqrt{\Heg}\frac{1}{\varepsilon})$ times walk operators is called to keep the part of initial state that matches  $1$-eigenvector and reduce the amplitude of other eigenvectors to $\varepsilon$. 
	However, heavy dependence on $\varepsilon$ influences not only the times of calling quantum interpolated walks operator but also the number of ancilla qubits. Unfortunately, phase estimation cannot help reduce them.
	In our work, these problems are all optimized here.

	The second method we apply is quantum fast-forwarding \cite{Apers2018QuantumFM}. It simulates the dynamics of classical random walk by using quantum walk operator, and quadratic speedup relative to the classical case can be achieved for searching marked vertices \cite{Ambainis2020QuadraticSF}.
	In general, any classical walk based on reversible Markov chain always falls into a stationary state after taking enough steps. The process can be simulated by using quantum fast-forwarding, which keeps the stationary state evolving like classical process while changing the ancilla qubits of other states, namely distinguishing 1- eigenvectors from other eigenvectors. The whole process can be used in place of phase estimation.
	
	Our algorithm combines interpolated walk and quantum fast-forwarding to obtain a qsampling algorithm with total complexity $\Theta(\sqrt{\Heg}\log(\frac{1}{\varepsilon}))$ and less ancilla qubits, which is currently the fastest algorithm in discrete-time case. For continue-time case, the state-of-the-art algorithm \cite{PhysRevA.102.022423} calls the same number of quantum operators but it requires more ancilla qubits.
	
	More importantly, the algorithm \cite{PhysRevA.102.022423} that satisfies the above complexity is only applicable to graphs with known $\pi_g$ for any vertex $g$.
	However, only regular graphs have known $\pi_g$ for a randomly chosen vertex $g$ without requiring any additional information since all vertices have equal probability, but non-regular graphs with unknown $\pi_g$ are more often encountered. Then it is particularly important to prepare the stationary state for non-regular graphs. Reverse search by the existing algorithm\cite{Krovi2015QuantumWC} will take total complexity $\Theta(\log(\pi_{min}^{-1})\sqrt{\Heg}\log(\frac{1}{\varepsilon}))$ to find $\pi_g$ by binary search, and the complexity increases at least additional $\log n$ compared with known $\pi_g$ case.
	We construct a new algorithm to determine the value of $\pi_g$, which reduces the additional cost to $\Theta(\log(\pi_{min}^{-1}/n))$. The additional cost becomes constant in most sparse graphs that satisfies $\pi_{min} = \Theta(1/n)$.
	Besides, a qsampling algorithm with constant success probability is not enough and amplitude amplification is applied to amplify success probability. A new reflection on stationary state with fewer ancilla qubits compared to reflection in \cite{Wocjan2008SpeedupVQ} is constructed  and  may have independent applications.
	
	Compared with the existing works, we build a qsampling algorithm that not only accelerates qsampling in non-regular graphs but also maintains the speed-up of existing quantum algorithms in regular graphs with fewer ancilla qubits. Actually, we provide a general time upper bound of qsmapling for all reversible Markov chains without conditions required. Our algorithm achieves the quadratic acceleration over the classical case in some applicable graphs, especially a series of sparse graphs that are difficult to quickly reach the stationary distribution in classical cases.
	
	Our main results are listed in the following theorem and summarized in \cref{tab:table0} and  \cref{tab:table1}.
	
	\begin{theorem}[Informal statement of main results]
		For an ergodic reversible Markov chain $P$ in $G(V,E)$, there exists an algorithm that achieves $\ket{\pi}$ with $1-\varepsilon$ success probability from an initial state of the form $\ket{g}$, where $g$ is any randomly selected vertex in graph $G$. 
		
		1. For non-regular graphs(the case with unknown $\pi_g$), the complexity is $\Theta(\log(\pi^{-1}_{min}/n)\sqrt{\Heg}\log(\varepsilon^{-1}))$ with $\Theta(\log(\pi^{-1}_{min}/n)(\log \sqrt{HT}+\log\log(\frac{1}{\varepsilon})))$ ancilla qubits.
		
		2. For regular graphs(the case with known $\pi_g$), the complexity is $\Theta(\sqrt{\Heg}\log(\varepsilon^{-1}))$ with $\Theta(\log( \sqrt{HT})+ \log\log(\frac{1}{\varepsilon}))$ ancilla qubits.	
	\end{theorem}
	\begin{table}[h!]
		\centering%	\begin{center}
		\resizebox{\linewidth}{!}{
			\begin{tabular}{c c| c c c c}
				\hline
				\multicolumn{2}{c}{\multirow{2}{*}{\textbf{Case}}} &\multirow{2}{*}{\textbf{Our result}}& \multicolumn{2}{c}{\textbf{Best previous quantum results}}&\multirow{2}{*}{\textbf{Best classical result\cite{https://doi.org/10.1112/jlms/s2-25.3.564}}}\\

				\multicolumn{2}{c}{} & & \textbf{Discrete-time\cite{Krovi2015QuantumWC}}& \textbf{Continue-time\cite{PhysRevA.102.022423}} & \\
				\hline
				%-----------------
				\multirow{3}{*}{\textbf{Complexity }} & \textbf{non-regular graphs} &  $\Theta(\log(\pi_{min}^{-1}/n)\log(\frac{1}{\varepsilon})\sqrt{\Heg})$ &$\Theta(\log(\pi_{min}^{-1})\frac{1}{\varepsilon}\sqrt{\Heg})$ &
				$\setminus$&$\Theta(\frac{1}{\delta}(\log(\pi_{min}^{-1})+\log(\frac{1}{\varepsilon})))$ \\
				\cline{2-6}
				\multirow{3}{*}{} & \textbf{sparse graphs} &  $\Theta(\log(\frac{1}{\varepsilon})\sqrt{\Heg})$ &$\Theta(\log(\pi_{min}^{-1})\frac{1}{\varepsilon}\sqrt{\Heg})$ &
				$\setminus$&$\Theta(\frac{1}{\delta}(\log(\pi_{min}^{-1})+\log(\frac{1}{\varepsilon})))$ \\
				\cline{2-6}
				
				\multirow{3}{*}{} & \textbf{regular graphs} &  $\Theta(\log(\frac{1}{\varepsilon})\sqrt{\Heg})$ & $\Theta(\frac{1}{\varepsilon}\sqrt{\Heg})$  &
				$\Theta(\log(\frac{1}{\varepsilon})\sqrt{\Heg})$&$\Theta(\frac{1}{\delta}(\log(\pi_{min}^{-1})+\log(\frac{1}{\varepsilon})))$ \\
				\hline
		\end{tabular}}
		\caption{Summary of our main results.}
		%	\end{center}	
		\label{tab:table0}
	\end{table}
	\begin{table}[H]
		\centering
		\resizebox{\linewidth}{!}{
			\begin{tabular}{c c| c c c}
				\cline{1-5}
				\multicolumn{2}{c}{\multirow{2}{*}{\textbf{Case}}} &\multirow{2}{*}{\textbf{Our result}}& \multicolumn{2}{c}{\textbf{Best previous quantum results}}\\
				%\cline{4-5}
				
				\multicolumn{2}{c}{} & & \textbf{Discrete-time\cite{Krovi2015QuantumWC}}& \textbf{Continue-time\cite{PhysRevA.102.022423}} \\
				\cline{1-5}
				%-----------------
				\multirow{2}{*}{\begin{tabular}{c} \textbf{non-regular graphs}\\{(unknown $\pi_g$)}\end{tabular}} & \textbf{Complexity} &  $\Theta(\log(\pi_{min}^{-1}/n)\sqrt{\Heg}\log(\frac{1}{\varepsilon}))$ &$\Theta(\log(\pi_{min}^{-1})\sqrt{\Heg}\frac{1}{\varepsilon})$ &
				$\setminus$\\
				\cline{2-5}
				
				\multirow{2}{*}{} & \textbf{Number of ancilla qubits} &$\Theta\left(\log(\pi_{min}^{-1}/n)( \log\sqrt{\Heg}+\log\log(\frac{1}{\varepsilon}) )\right)$	& $\Theta\left(\log(\pi_{min}^{-1})(\log\sqrt{\Heg}+\log\frac{1}{\varepsilon})\right)$ 	&$\setminus$	\\
				\cline{1-5}
				\multirow{2}{*}{\begin{tabular}{c}\textbf{regular graphs}\\(known $\pi_g$)\end{tabular}}
				& \textbf{Complexity} &  $\Theta(\sqrt{\Heg}\log(\frac{1}{\varepsilon}))$ & $\Theta(\sqrt{\Heg}\frac{1}{\varepsilon})$  &
				$\Theta(\sqrt{\Heg}\log(\frac{1}{\varepsilon}))$\\
				\cline{2-5}
				
				\multirow{2}{*}{} & \textbf{Number of ancilla qubits} &$\Theta( \log\sqrt{\Heg}+\log\log(\frac{1}{\varepsilon}) )$	& $\Theta(\log(\sqrt{\Heg}+\log\frac{1}{\varepsilon}))$ 	&$\Theta(\log\sqrt{\Heg}\cdot\log(\frac{1}{\varepsilon}))  $	
				\\
				\cline{1-5}
		\end{tabular}}	
		% \end{center}
		\caption{Summary of our main results. Here $\delta$ denotes the eigenvalue gap of Markov chain $P$, $\pi_{min}\coloneqq \min_{g \in V}\pi_g$ denotes the smallest probability component in stationary state, $\Heg$ is the classical hitting time bound of $G$ for any vertex in $V$, and $\varepsilon$ denotes the error. Formal results can be found in \cref{thm:qsampling with known pm} and \cref{thm:qsampling with bound on p}.}\label{tab:table1}
	\end{table}
	\begin{remark}
		Although the stationary state with unknown $\pi_g$ is not prepared in continuous-time case \cite{PhysRevA.102.022423}, we can give a continuous case result according to \cite{PhysRevA.102.022423}, which can be prepared with  $\Theta(\log(\pi_{min}^{-1})\sqrt{\Heg}\log(1/\varepsilon))$ complexity and $\Theta(\log(\pi_{min}^{-1})\log\sqrt{\Heg}\log(\frac{1}{\varepsilon})) $ ancilla qubits. Our discrete qsampling algorithm is faster than continuous-time case with less ancilla qubits in unknown $\pi_g$ case.
	\end{remark}
	
	The paper is organized as follows. 
Preliminaries and our technical overview are provided in \cref{sect:Technical overview} firstly. Second we introduce quantum fast-forwarding to improve the reflection on $\ket{\pi}$ in \cref{sect:reflection pi algorithms}, which reduces the number of ancilla qubits required.
	Next we combine quantum fast-forwarding with quantum interpolated walks. A state that has a constant overlap with $\ket{\pi}$ is prepared for both regular graphs and non-regular graphs in \cref{sect:Search algorithms}.  
	At last, complexity analysis and scope of usage are discussed in \cref{sect:Discussion}, where the paper is concluded.  
	%\vspace{-0.5cm}
	
	\section{Technical overview}\label{sect:Technical overview}
	\subsection{Preliminaries}\label{sect:Preliminaries}
	\hspace{1.2em}
	Here we describe some details on Markov chains firstly.	Consider a Markov chain P on graph $G(V, E)$ with finite state space of size $ n=\left | V \right |$, where $V$ and $E$ represent vertex set and edge set of $G$ respectively. Let $P=(p_{xy})_{x,y \in V}$ be transition matrix of Markov chain P, where $p_{xy}$ denotes the transition probability from $x$ to $y$ for arbitrary vertex $x,y \in V$. A Markov chain is ergodic if each state can reach any other state after enough steps and the length of each directed loop has the greatest common factor of $1$. Ergodic chain has a unique $1$-eigenvector, which is the stationary distribution represented by $\pi={( \pi_x ) }_{x\in V}$. A Markov chain is reversible if it satisfies $\pi_xp_{xy} = \pi_yp_{yx}$.
	
	Let $\lambda_0$ denote the eigenvalue of reversible $P$ that equals to $1$, and the remaining eigenvalues $\lambda_j$ lies in $[-1,1)$ for $j = 1, \dotsc, n-1$ in  non-increasing order. Let $\delta\coloneqq 1 - \lambda_1$ denote the eigenvalue gap of $P$.
	To satisfy all eigenvalues of $P$ lies in $[0,1]$ below, we introduce lazy walk $P^{lazy}=(I+P)/2$ and replace $P$ with $P^{lazy}$ for the unsatisfied case.
	Without loss of generality, the transformation affects complexity only by adding a constant factor $2$ which is shown in \cref{rem:lazy}. 
	
	Sampling in reversible Markov chain is to prepare stationary distribution $\pi$. Similarly, qsampling is to prepare stationary state $\ket{\pi} \coloneqq \sum_{g \in V}\sqrt{\pi_g}\ket{x}\in \Lambda_V$, where $\Lambda_V \coloneqq \spn\{\ket{x}: x \in V\}$.
	
	Quantum walks take place on the extended space $\Lambda_V \x \Lambda_V$. For arbitrary initial probability distribution $\sigma$ over $V$, the initial state is defined as $\ket{\sigma}\ket{\bar{0}}$,  where the state in the second register is some fixed initialization state in $\Lambda_V$.
	The quantum walk operator corresponding to transition matrix $P$ can be expressed as $$W \coloneqq V^{\dagger} \cdot S \cdot V \cdot R_{\bar{0}}.$$ $V$ is a unitary operator that derived from the transition matrix $P$ and satisfies
	$$V \ket{x} \ket{\bar{0}}\coloneqq \ket{x} \sum_{y \in X} \sqrt{P_{xy}} \ket{y}.$$
	The shift operator $S$ and the reflection operator $R_{\bar{0}}$ are expressed as
	$$	\ket{x}\ket{y}\mapsto S\ket{x}\ket{y} \coloneqq
	\begin{cases}
		\ket{y}\ket{x}, & \text{if } (x,y) \in E, \\
		\ket{x}\ket{y}, & \text{otherwise.}
	\end{cases}$$
	$$	\ket{x}\ket{y}\mapsto R_{\bar{0}}\ket{x}\ket{y} \coloneqq I\x \left( 2\ket{\bar{0}}\bra{\bar{0}} - I\right)\ket{x}\ket{y}.$$

	Consider a graph with some marked vertices and let $M$ be the  marked vertex set. In order to find any vertex of $M$ faster, interpolation is introduced into Markov chains \cite{Krovi2015QuantumWC, Krovi2010AdiabaticCA}. The interpolated Markov chain $P(s)$ can be expressed as
	{\tiny }		$$P(s) \coloneqq (1-s)P + sP',$$
	where $0\le s \le 1 $, $P'$ is a modified version of $P$ that all outgoing transformations from marked vertices are replaced by self-loops, and corresponding $W(s)$ are defined as $W(P(s))$ as follows:
	\begin{equation}\label{eq:ws}
		W(s) = V(s)^{\dagger}(s)\cdot S\cdot V(s) \cdot R_{\bar{0}}, \quad where \quad V(s) \ket{x} \ket{\bar{0}}\coloneqq \ket{x} \sum_{y \in X} \sqrt{P(s)_{xy}} \ket{y}.
	\end{equation}
	Let $D$ denote the discriminant matrix of Markov chain P as
		$D\coloneqq \sqrt{P \circ P^{\mathsf{T}}}$,
		where the “$\circ$” and the square root are computed elementwise and we have $D\ket{\pi} = \ket{\pi}$.
		$D(s)$ is defined as $D(P(s))$ similarly. In classical random walk based on Markov chain P, the time to hit any marked vertex in $M$ from the initial distribution $\bar{\pi}$ can be represented by  hitting time as
		%	The complexity of quantum interpolated walks is related to classical hitting times.
		$$	\Heg(M) \coloneqq \sum_{k=1}^{n-|M|}\frac{|\langle v_k'| \bar{\pi} \rangle|^2 }{1-\lambda_k' },$$
	as the classical hitting time of Markov chain $P$ relative to the marked set $M$ in random walk, where $\lambda_k'$ and $\ket{v_k'}$ are eigenvalues and the corresponding eigenvectors of the discriminant matrix $D = D(P')$ in nondecreasing order, and 
	\begin{equation}\label{eq:barpi}
		\ket{\bar{\pi}} \coloneqq \frac{1}{\sqrt{1-\pi_M}}\sum_{x \notin M}\sqrt{\pi_x}\ket{x}.
	\end{equation}
	In order to hold for any $g \in V$, $\Heg$ will represent $\max_{x \in V} \Heg(x)$.
	
	\subsection{Main algorithm framework}\label{sect:T}
\begin{figure}[h!]
		\centering 
		\includegraphics[width=1\linewidth]{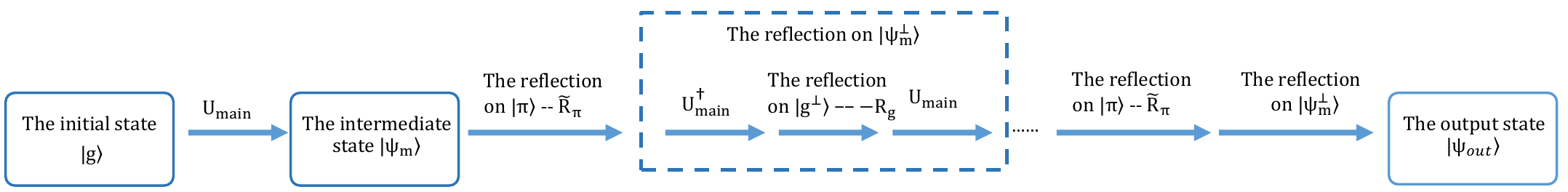}
		\caption{Main algorithm framework}
		\label{fig:a00}
	\end{figure}

	\hspace{1.2em}
We determine the value of $\pi_g$ firstly to prepare qsampling from any reversible Markov chain with the initial state $\ket{g}$. The corresponding process can be found in \cref{sect:unknown pm}, where $\pi_g$ is determined faster by narrowing its possible range. Then the rest can be seen as preparing stationary state $\ket{\pi}$ with known $\pi_g$ and the corresponding algorithm flow refers to \cref{fig:a00}.
%A new algorithm to prepare stationary state $\ket{\pi}$ with known $\pi_g$ will be constructed by using the reflection on $\ket{\pi}$ and reflection on intermediate state to implement amplitude amplification, where intermediate state can be obtained by quantum fast-forwarding processing for the initial state. 
	
	The first phase is to construct the intermediate state $\ket{\psi_m}$ that has a constant overlap with the target state $\ket{\pi}$. In this part we apply quantum fast-forwarding algorithm into interpolated walk by constructing operator according to initial state $\ket{g}$. 
	An appropriate interpolation parameter $s$ is chosen such that both the initial state and the stationary state have constant overlap with the 1-eigenvector of quantum interpolated walk operator. 
	The method of quantum fast-forwarding is used to distinguish $1$-eigenvector from other eigenvectors by using quantum walk operator to simulate the dynamics of classical random walk, which reduces the dependence on $\pi_{min}$ and on $\varepsilon$. Besides, the number of ancilla qubits required is reduced as well. And then a state that has constant overlap with the $\ket{\pi}$ will be prepared and we call it intermediate state $\ket{\psi_m}$.
	
	In the second phase, we improve the reflection on $\ket{\pi}$ by quantum fast-forwarding. By combining reflection on intermediate state $\ket{\psi_m}$ and amplitude amplification method, we can $\varepsilon$-close the expected stationary state $\ket{\pi}$ and output it.
	
	The above process consists of three unitary operator: $U_{main}(U^{\dagger}_{main})$, $\tilde{R}_{\pi}$ and $R_g$. Our work consists mainly of the specific structure of $U_{main}$ ($i.e.$ the preparation of intermediate state $\ket{\psi_m}$) and $\tilde{R}_{\pi}$ ($i.e.$ the approximate reflection on $\ket{\pi}$) since the specific structure of $R_g$ ($i.e.$ the reflection on the initial state $g$) is trivial. We first introduce the approximate reflection on $\ket{\pi}$ in \cref{sect:reflection pi algorithms}, and then introduce the preparation of $\ket{\psi_m}$ in \cref{sect:Search algorithms} for technical reasons.
	
	\section{Construction of quantum reflection on $\ket{\pi}$} \label{sect:reflection pi algorithms}
	\hspace{1.2em}
	The state that has constant overlap with stationary state is assumed to be obtained here, which will be prepared specifically in \cref{sect:Search algorithms}. Amplitude amplification is introduced to improve the qsampling success probability from constant to $1 - \varepsilon$, and the reflection on the target state $\ket{\pi}$ is an indispensable part.
	
	There exists many approaches to prepare the reflection on stationary state \cite{Wocjan2008SpeedupVQ, doi:10.1137/1.9781611975994.12, Orsucci2018fasterquantummixing} by quantum walks and they all use phase estimation. If the state to be detected is $\ket{\pi}$, namely the 1-eigenvector of the walk operator, phase estimation remains the ancilla qubits unchanged and add a phase; otherwise, the ancilla qubits is changed according to the eigenvalue.
	
	Here we introduce an approximate reflection using quantum fast-forwarding instead of phase estimation.
	Quantum fast-forwarding algorithm simulates $t$ steps classical random walk action by $\sqrt{t}$ steps quantum walks, which is used to solve the search marked vertices problem \cite{Ambainis2020QuadraticSF}. 
	Invoking quantum fast-forwarding here improves the reflection on $\ket{\pi}$, greatly reducing the number of ancilla qubits, while maintaining the number of calls to the walk operator. We believe this subroutine will help solving other problems as well, such as search problem in \cite{doi:10.1137/090745854}.

	\begin{lemma}[Quantum fast-forwarding~\cite{Apers2018QuantumFM}]\label{lem:quantum fast-forwarding}
		For any reversible Markov chain $P$ and corresponding operator $W$ on state space V, any initial state $\ket{\psi}\ket{\bar{0}} \in \spn \{\ket{x}\ket{\bar{0}}: x \in V\}$, $\varepsilon_1 \in (0,1)$, there exists a quantum fast-forwarding algorithm QFF$(W, t, \varepsilon_{1})$, which invokes controlled-$W$ operator $\Gamma \coloneqq \Theta( \sqrt{t\log(\varepsilon_{1}^{-1})}) $ times, acts on  $2\left \lceil \log n \right \rceil +\tau$ qubits, and outputs a state $\varepsilon_{1}$-close to $$(D^{t}\ket{\psi})\ket{\bar{0}}\ket{0^{\tau}} + \ket{ \Upsilon },$$
		where the third register with initial state $\ket{0^{\tau}}\coloneqq\ket{0}^{\x{\tau} }\in \spn\lbrace \ket{0}, \ket{1}, \dots, \ket{2^{\tau}-1}\rbrace $ is an ancilla register,  $\ket{ \Upsilon }$ is any state of the form perpendicular to  $\ket{\bar{0}0^{\tau}}$ in the last two registers, $\tau = \lceil \log\Gamma\rceil$.
	\end{lemma}
	The dynamics of discriminant matrix $D$ is simulated by the unitary operator $W_{\tau}$, which acts as follows:
	\begin{equation} \label{eq:close}
		\left \| \Pi_{\bar{0}0}W_{\tau}\ket{\psi}\ket{\bar{0}0^{\tau}} - (D^{t}\ket{\psi})\ket{\bar{0}0^{\tau}}  \right \| \le \varepsilon_{1},
	\end{equation}
	where $\Pi_{\bar{0}0} = I \x (\ket{\bar{0}}\bra{\bar{0}})\x(\ket{0^{\tau}}\bra{0^{\tau}})$.	The detailed construction of $W_{\tau} = V_{q}^{\dagger} W_{ctrl} V_{q}$ can be described as: 
	\begin{equation}\label{eq:vqwctrl}
		V_q \ket{\psi,\bar{0}}\ket{0^{\tau}}= \sum_{l=0}^{2^{\tau}-1} \sqrt{\frac{p_l}{1-(\sum_{x = {2^{\tau}}}^{t}p_x)}} \ket{\psi,\bar{0}}\ket{l},\ \ W_{ctrl} = \sum_{l=0}^{2^{\tau}-1} W^l \x \ket{l}\bra{l},
	\end{equation}and the detailed description of $p_l$ can be found in Equation (6) of \cite{Apers2018QuantumFM}. 
	
	We construct the circuit of quantum fast-forwarding algorithm as below.
	\begin{figure}[h!]
		\centering
		\includegraphics[width=0.6\linewidth]{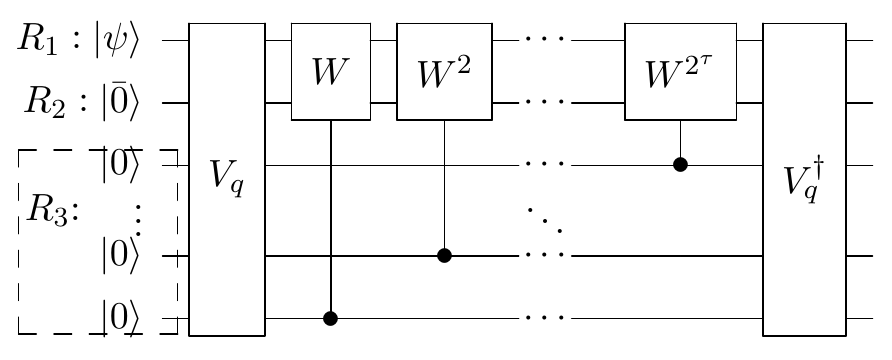}
		\caption{Circuit $W_{\tau}$ of quantum fast-forwarding algorithm}
		\label{fig:QFFcircuit}
	\end{figure}
	\begin{remark}\label{rem:qff}
		 The relationship between the number of ancilla qubits $\mathrm{(\tau)}$ and the number of controlled walk operator $\mathrm{(\Gamma)}$ is shown in \cref{fig:QFFcircuit}. Therefore, an upper bound on the number of steps is required to determine the number of ancilla qubits before executing the algorithm.
	\end{remark}
	Next, we use quantum fast-forwarding to construct the approximate reflection $\tilde{R}_{\pi}$ on $\ket{\pi}$, which changes the phase of $\ket{\pi}$ and keeps the phase of any other state roughly constant. Let $\ket{\psi_{0}}$ denote the $1$-eigenvector of $D$, $\ket{\psi_{j}}$ and $cos(\theta_j)$
	denote the remaining eigenvectors and eigenvalues (in non-increasing order) for $j=1,\ldots,n-1$, and we have the following result. %For any state $\ket{\psi} = \sum^{n-1}_{j=0}\alpha_j\ket{\psi_j}$, 
	\begin{theorem}
		$P$ is an ergodic reversible Markov chain and $\delta$ denotes the corresponding eigenvalue gap.
		For any $\varepsilon_{2} \in (0,1)$, there exists a quantum circuit $\tilde{R}_{\pi}$ that invokes controlled-W operator $\Theta(\frac{1}{\sqrt{\delta}}\log(\frac{1}{\varepsilon_{2}})) $ times, acts on $2 \lceil{\log n \rceil} + \Theta (\log(\frac{1}{\sqrt{\delta}})+\log(\log(\frac{1}{\varepsilon_{2}})))$  qubits, and satisfies the following properties:
		\begin{equation} \label{eq:UP0close}
			\|  ( \tilde{R}_{\pi}+I)  \ket{\psi_{0}}\ket{\bar{0}0^{\tau}0} \|^2 =0
		\end{equation}
		\begin{equation} \label{eq:UPkclose}
			\|  ( \tilde{R}_{\pi}-I)  \ket{\psi_{j}}\ket{\bar{0}0^{\tau}0} \|^2 \le   \varepsilon_2
		\end{equation}
		for $j=1,\ldots,d-1$.
		\label{thm:2}
	\end{theorem}
	
\begin{figure}[h!]
		\centering 
		\includegraphics[width=1\linewidth]{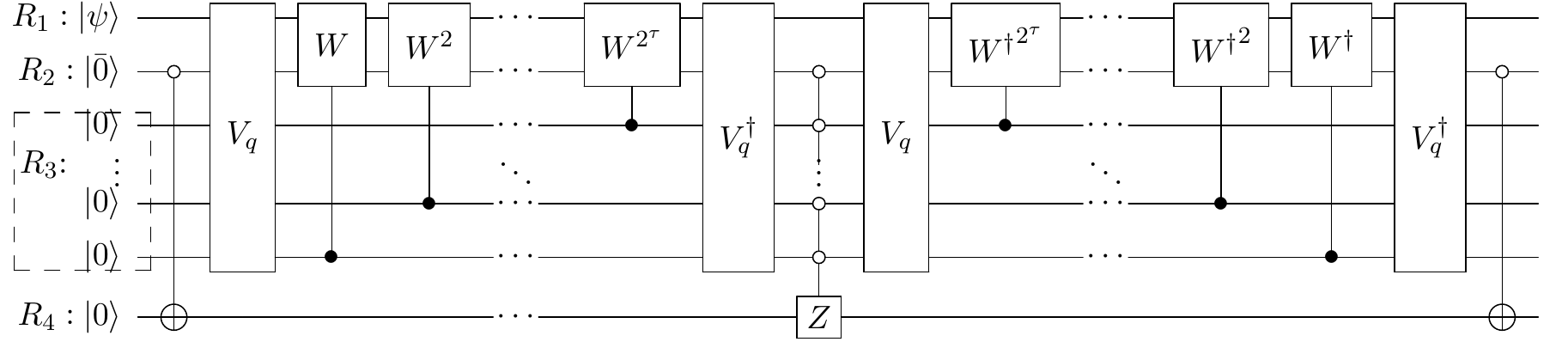}
		\caption{Quantum circuit $\tilde{R}_{\pi}$ as reflection on $\ket{\pi}$}
		\label{fig:reflectioncircuit}
	\end{figure}

	%	\begin{proof}
	For simplicity, we define some operator symbols. For any operator $A$, let $A_{x}$ represent that the $x$th register is applied with $A$ and the other registers are applied with $I$. $cA_{x,y}$ represents the controlled-$A$ gate, where $x$ is the register of control qubit and $y$ is the register of target qubit. $A$ is applied when the control qubit is $\ket{1}$, $ccA_{x,y}$ represents the controlled-$A$ gate, $A$ is applied when the control qubit is $\ket{0}$, and the qubits in other registers is applied by $I$.
	First, we define $$\tilde{R}_{\pi} \coloneqq ccX_{2,4}( W_{\tau}^{\dagger} \x I)ccZ_{23,4}( W_{\tau} \x I)ccX_{2,4},$$ the corresponding circuit\footnote{For simplicity, we set the second register of the initial state $\ket{\bar{0}}$ as $\ket{0^{2^{\left \lceil \log{n} \right \rceil}}}$ in the circuit. Only all $2^{\left \lceil {\log{n}} \right \rceil}$ qubits in control register are $0$, the controlled unitary operator is applied.} can be seen in \cref{fig:reflectioncircuit}.
	There are four registers in circuit and we set them as $\Reg_1\Reg_2\Reg_3\Reg_4$. The first three registers $\Reg_1\Reg_2\Reg_3$ is the same as \cref{fig:QFFcircuit} and the fourth register $\Reg_4$ is single-qubit with initial state $\ket{0} \in \spn\{ \ket{0}, \ket{1}\}$.  $\Reg_3$ and $\Reg_4$ are both ancilla registers.	
	
	Since $\ket{\psi_0}$ is the $1$-eigenvector of $W$, if the input quantum state is $\ket{\psi_0}\ket{\bar{0}0^{\tau}0}$, from \cref{fig:reflectioncircuit} we have
	\begin{align}
		& \| ( \tilde{R}_{\pi}+I) \ket{\psi_{0}}\ket{\bar{0}0^{\tau}0} \| \tag*{ }\\
		&=  \| ccX_{2,4}( W_{\tau}^{\dagger} \x I)(ccZ_{23,4}+I)( W_{\tau} \x I)ccX_{2,4}\ket{\psi_{0}}\ket{\bar{0}0^{\tau}0} \| \tag*{ }\\
		&=  \| (2I-2\Pi_{\bar{0}0} \x \ket{1}\bra{1} ) ( W_{\tau} \x I)\ket{\psi_{0}}\ket{\bar{0}0^{\tau}}\ket{1} \| \tag*{by norm invariance after unitary operator}\\
		&=  2\| (I-\Pi_{\bar{0}0} )\ket{\psi_{0}}\ket{\bar{0}0^{\tau}}\| = 0,\tag*{ }
	\end{align}
	which means $\tilde{R}_{\pi}$ can change the phase of $\ket{\psi_0}$ perfectly, and the remaining goal is to keep the other states unchanged.
	
	When $j \ne 0$, we have
	\begin{align}
		&\|( \tilde{R}_{\pi}-I) \ket{\psi_{j}}\ket{\bar{0}0^{\tau}0}  \|\tag*{ }\\
		&= \| ccX_{2,4}( W_{\tau}^{\dagger} \x I) (ccZ_{23,4}-I ) ( W_{\tau} \x I)ccX_{2,4}\ket{\psi_{j}}\ket{\bar{0}0^{\tau}0} \| \tag*{ }\\
		&= \| (-2\Pi_{\bar{0}0} \x \ket{1}\bra{1} ) ( W_{\tau} \x I)\ket{\psi_{j}}\ket{\bar{0}0^{\tau}}\ket{1} \|\tag*{by norm invariance after unitary operator} \\
		&\le 2\| \Pi_{\bar{0}0} W_{\tau}\ket{\psi_{j}}\ket{\bar{0}0^{\tau}}\| \tag*{by trigonometric inequality}\\
		&\le 2\| D^{t}\ket{\psi_{j}}\ket{\bar{0}}\ket{0}^{\x \tau}\|  +  2\| \Pi_{\bar{0}0}W_{\tau}\ket{\psi_j}\ket{\bar{0}}\ket{0}^{\x \tau} - D^{t}\ket{\psi_j}\ket{\bar{0}}\ket{0}^{\x \tau}  \|.\tag*{ }
	\end{align}	
	Since $\Gamma = \Theta( \sqrt{t\log(\varepsilon_{1}^{-1})})$, let $t = \log(\frac{4}{\varepsilon_{2}})\frac{2}{\delta}$ and $\varepsilon_{1} = \frac{\varepsilon_{2}}{4}$, and then we obtain
	\begin{equation}
		\cos^{t}\varphi_j \le \cos^{t}\varphi_1= (1 - \delta)^t =  e^{-t \log(\frac{1}{1-\delta})} = \left[e^{- \log(\frac{4}{\varepsilon_2})} \right]^{\frac{2}{\delta}\log(\frac{1}{1-\delta})} = (\frac{\varepsilon_2}{4})^{\frac{2}{\delta}\log(\frac{1}{1-\delta})} \le \frac{\varepsilon_2}{4},
		\label{eq:ccos}
	\end{equation}  
	where the last inequality is due to
	$$\frac{2}{\delta}\log( \frac{1}{1-\delta}) = \frac{2}{\delta}\log( 1 + \frac{\delta}{1-\delta})  = \frac{2}{\delta}( \frac{\delta}{1-\delta}-\frac{1}{2}( \frac{\delta}{1-\delta}) ^{2}+\ldots+( -1) ^{n-1}\frac{1}{n}( \frac{\delta}{1-\delta}) ^n+ \ldots ) \ge 1. $$
	Then we have
	\begin{equation} \label{eq:ccc}
		\|D^{t}\ket{\psi_{j}}\ket{\bar{0}} \|= 	\|\cos^{t}( \theta_j) \ket{\psi_{j}}\ket{\bar{0}} \|=  \cos^{t}( \theta_j) \le \frac{\varepsilon_2}{4}. 
	\end{equation}
	Combine \cref{eq:close} and \cref{eq:ccc}, then we obtain \cref{eq:UPkclose}.
	
	Now we analyze the complexity of $\tilde{R}_{\pi}$. From \cref{fig:reflectioncircuit}, the circuit of $\tilde{R}_{\pi}$ invokes controlled-$W$ $2^{\tau + 1}$ times and $\tau$ ancilla qubits, where
	$$\Gamma = \Theta( \sqrt{t\log(\varepsilon_{1}^{-1})}) = \Theta( \log(\frac{1}{\varepsilon_2})\frac{1}{\sqrt{\delta}}),\quad \tau = \lceil \log\Gamma\rceil= \Theta(  \log(\frac{1}{\sqrt{\delta}}) + \log\log(\frac{1}{\varepsilon_{2}}) ).$$
	%	\end{proof}
	
Compared with the reflection on $\ket{\pi}$ based on phase estimation \cite{Wocjan2008SpeedupVQ}, with $\Theta(\log(\frac{1}{\sqrt{\delta(P)}})\cdot\log(\frac{1}{\varepsilon_{2}}))$ ancilla qubits, the number of our ancilla qubits $ \Theta(\log(\frac{1}{\sqrt{\delta}})+\log\log(\frac{1}{\varepsilon_{2}}))$ is much smaller. 
Therefore, our approximate reflection greatly reduces the number of ancilla qubits, while maintaining the number of walk operator invoked. 

	\begin{remark}
		\label{rem:3}
		 The effect of $\tilde{R}_{\pi}$ in \cref{fig:reflectioncircuit} is the same as $( W_{\tau}^{\dagger} \x I)ccZ_{23,4}( W_{\tau} \x I)$ when we start from the initial state $\ket{\cdots}\ket{\bar{0}}\ket{\cdots}$. It seems unnecessary to add two $ccX_{2,4}$ and here we discuss their usefulness.
		
		The output remains unchanged  when the first two registers of input is $\ket{\pi}\ket{\bar{0}}$ according to \cref{thm:2}, which means the second register of output is $\ket{\bar{0}}$.
		Thus if the output is expressed as $\ket{\cdots}\ket{\bar{0}^{\perp}}\ket{\cdots}$, the  first register of input must be $\ket{\pi^{\perp}}$, $i.e.$ the state should not be added phase in the circuit of reflection on $\ket{\pi}$, and the two $ccX_{2,4}$ in the circuit make sure that the phase of these input state is unchanged.
	\end{remark}
	
	\section{Preparation of stationary state $\ket{\pi}$} \label{sect:Search algorithms}
	\hspace{1.2em}
	Let $\pi_g$ denote the square of overlap between the single-vertex initial state $\ket{g}$ and stationary state $\ket{\pi}$.
	Here we show intermediate state $\ket{\psi_m}$ that has constant overlap with stationary state $\ket{\pi}$ can be prepared with complexity $\Theta(\sqrt{\Heg}\log\frac{1}{\varepsilon})$ by introducing quantum interpolated walks to quantum fast-forwarding algorithm with known $\pi_g$, and amplitude amplification is used to amplify success probability.
	
	Then a new algorithm to prepare the stationary state with unknown $\pi_g$ is constructed, which reduces the cost of determining $\pi_g$ from $\log(\pi_{min}^{-1})$ to $\log(\pi^{-1}_{min}/n)$. The additional cost to determine $\pi_g$ will reduce to constant in a series sparse graphs.
	\subsection{The case with known $\pi_g$}\label{sect:known pm}
	Assume the value of $\pi_g$ has been determined, such as regular graphs, then the intermediate state $\ket{\psi_{m}}$ that has constant overlap with stationary states is prepared from the single-vertex initial state $\ket{g}$ as follows.
	\begin{theorem}\label{thm:qsampling with known pm}
		For an ergodic reversible Markov chain $P$ in $G(V,E)$, \cref{alg:qis} achieves $\ket{\pi}$ with constant success probability, and the corresponding complexity is $\Theta(\log(\varepsilon^{-1})\sqrt{\Heg})$, where $\varepsilon$ is error.
	\end{theorem}
	%	\begin{proof}	
	\begin{algorithm}[h!]	
		\caption{The algorithm of preparing $\ket{\psi_{m}}$ with known $\pi_g$}
		\label{alg:qis}
		\textbf{Input}: $g$, an randomly chosen vertex in $V$; $\pi_g$, the probability of vertex $g$ in the stationary distribution $\pi$; $\varepsilon$, the expected error; $\Gamma$, a number satisfies that $\Gamma = \Theta({\log(\varepsilon^{-1})\sqrt{\Heg} })$.\\
		\textbf{Output}: $\ket{\psi_{m}}$, a state that has constant overlap with $\ket{\pi}$.
		
		\begin{algorithmic}[1]
			\STATE Set $\tau = \lceil  \log \Gamma \rceil$, prepare the initial state as $\ket{g}\ket{\bar{0}}\ket{0^{\tau}}\ket{0}\ket{0}$ on $\Reg_1\Reg_2\Reg_3\Reg_4\Reg_5$, where the fourth registers are the same as \cref{thm:2} and the fifth register $\Reg_5$ is a single-qubit register as $\spn\{ \ket{0}, \ket{1}\}$.	
			\STATE Apply $U_{main} \coloneqq ccX_{2,4}X_4(W(s)_{\tau}^{\dagger} \x I) ccX_{234,5} ( W(s)_{\tau} \x I)X_4ccX_{2,4}$ to the initial state, where $W(s)_{\tau}  = V_{q}^{\dagger}( \sum_{l=0}^{2^{\tau}-1} W(s)^l \x \ket{l}\bra{l})V_{q}$, $M =\{g\}$, $s = 1-\frac{\pi_g}{1-\pi_g}$.
			\STATE Output the state in $\Reg_1$ as $\ket{\psi_{m}}$.
		\end{algorithmic}
	\end{algorithm}
	
	\textbf{The success probability of algorithm with known $\pi_g$}

	Here we analyze the success probability firstly. For initial state $\ket{g}$ and target state $\ket{\pi}$, let $p_{succ}$ denote the success probability and we have\footnote{Here the effect of $ccX_{2,4}$ and $X_4$ is the same as \cref{rem:3}, to distinguish $\ket{\cdots}\ket{\bar{0}}\ket{\cdots}$ from $\ket{\cdots}\ket{\bar{0}^{\perp}}\ket{\cdots}$.}
	\begin{align}
		\sqrt{p_{succ}}		
		&= \| \ket{\pi}\bra{\pi}_{1}U_{main}(\ket{g}\ket{\bar{0}0^{\tau}00}) \|\tag*{}\\
		&= \|\ket{\pi}\bra{\pi}_{1} ccX_{2,4}X_4(W(s)_{\tau}^{\dagger} \x I) ccX_{234,5} ( W(s)_{\tau} \x I) X_4ccX_{2,4}\ket{g}  \ket{\bar{0}0^{\tau}00}\|  \tag*{} \\
		&\ge \|\ket{\pi}\bra{\pi}_{1} \ket{1}\bra{1}_{5}( W(s)_{\tau}^{\dagger} \x I) ccX_{234,5} ( W(s)_{\tau} \x I) \ket{g} \ket{\bar{0}0^{\tau}00}\|,  
		\label{eq:psuccess} \end{align}	where the detailed construction of $U_{main}$ can be seen in \cref{eq:ws} and \cref{eq:vqwctrl}. 
	Define \begin{equation}
		\ket{v_0(s)} \coloneqq \sqrt{\frac{(1-s)(1-\pi_g)}{1-s(1-\pi_g)}} \ket{\bar{\pi}} + \sqrt{\frac{\pi_g}{1-s(1-\pi_g)}} \ket{g},
		\label{eq:v0}	\  where \ \ket{\bar{\pi}} = \frac{1}{\sqrt{1-\pi_g}}( \ket{\pi} - \sqrt{\pi_g} \ket{g}),
	\end{equation}
	and we have $W(s)\ket{v_0(s)} = \ket{v_0(s)}$, $\ket{v_0(s)}$  is eigenvector of $W(s)$ and the corresponding eigenvalue is 1. Let $\ket{v_j(s)}$ denote the remaining eigenvectors (the corresponding eigenvalues are non-increasing) of $W(s)$ for $j=1,\ldots,n-1$. Thus initial state $\ket{g}\ket{\bar{0}}$ of the first two registers can be expressed as
	\begin{equation}\label{eq:g}
		\ket{g} \ket{\bar{0}} = \beta_0 \ket{v_0(s)} \ket{\bar{0}} + \sum_{k=1}^{n-1} \beta_k \ket{v_k(s)} \ket{\bar{0}}, \quad\textrm{where}\  \	\beta_0(s) \coloneqq \langle{v_0(s)}|{g}\rangle, \ \ \beta_k(s) \coloneqq\langle{v_k(s)}|{g}\rangle.
	\end{equation}	
	By triangle inequality and 	\cref{eq:psuccess} we have
	\begin{align}
		\sqrt{p_{succ}}	
		&\ge \|(\ket{\pi}\bra{\pi}_{1} \ket{1}\bra{1}_{5})( W(s)_{\tau}^{\dagger} \x I) ccX_{234,5} ( W(s)_{\tau} \x I)\beta_0(s) \ket{v_0(s)}  \ket{\bar{0}0^{\tau}00}\|  \tag*{} \\
		&- \|(\ket{\pi}\bra{\pi}_{1} \ket{1}\bra{1}_{5})( W(s)_{\tau}^{\dagger} \x I) ccX_{234,5} ( W(s)_{\tau} \x I)(\sum_{k=1}^{n-1} \beta_k(s) \ket{v_k(s)}) \ket{\bar{0}0^{\tau}00}\| \tag*{} \\
		&\ge | \beta_0(s) \left \langle \pi | v_0(s)\right \rangle  | - \|(\ket{1}\bra{1}_{5})( W(s)_{\tau}^{\dagger} \x I) ccX_{234,5} ( W(s)_{\tau} \x I)(\sum_{k=1}^{n-1} \beta_k(s) \ket{v_k(s)}) \ket{\bar{0}0^{\tau}00}\| \tag*{} \\
		&\ge | \beta_0(s) \left \langle \pi | v_0(s)\right \rangle  | - \| \Pi_{\bar{0}0} ( W(s)_{\tau} \x I)(\sum_{k=1}^{n-1} \beta_k(s) \ket{v_k(s)}) \ket{\bar{0}0^{\tau}00}\|.	
		\label{eq:sss}
	\end{align}
	The last inequality holds because the fifth register changes from $\ket{0}$ to $\ket{1}$ only if the state before $ccX_{234,5}$ is expressed as $\ket{\cdots}\ket{\bar{0}0^{\tau}00}$, and the detailed proof is listed as below.
	\begin{align}&\|(\ket{\pi}\bra{\pi}_{1} \ket{1}\bra{1}_{5})( W(s)_{\tau}^{\dagger} \x I) ccX_{234,5} ( W(s)_{\tau} \x I)(\sum_{k=1}^{n-1} \beta_k(s) \ket{v_k(s)}) \ket{\bar{0}0^{\tau}00}\| \tag*{} \\
		&\le\|(\ket{1}\bra{1}_{5})( W(s)_{\tau}^{\dagger} \x I) ccX_{234,5} ( W(s)_{\tau} \x I)(\sum_{k=1}^{n-1} \beta_k(s) \ket{v_k(s)}) \ket{\bar{0}0^{\tau}00}\| \tag*{} \\
		&=\|( W(s)_{\tau}^{\dagger} \x I)(\ket{1}\bra{1}_{5}) ccX_{234,5} ( W(s)_{\tau} \x I)(\sum_{k=1}^{n-1} \beta_k(s) \ket{v_k(s)}) \ket{\bar{0}0^{\tau}00}\| \tag*{by norm invariance after unitary operator} \\
		&	=\|(\ket{1}\bra{1}_{5}) ccX_{234,5} ( W(s)_{\tau} \x I)(\sum_{k=1}^{n-1} \beta_k(s) \ket{v_k(s)}) \ket{\bar{0}0^{\tau}00}\| \tag*{} \\
		&=\|(\ket{\bar{0}}\bra{\bar{0}}_{2})(\ket{0^\tau}\bra{0^\tau}_{3})(\ket{0^\tau}\bra{0}_{4})  ( W(s)_{\tau} \x I)(\sum_{k=1}^{n-1} \beta_k(s) \ket{v_k(s)}) \ket{\bar{0}0^{\tau}00}\| \tag*{} \\
		&\le\|(\ket{\bar{0}}\bra{\bar{0}}_{2})(\ket{0^\tau}\bra{0^\tau}_{3}) ( W(s)_{\tau} \x I)(\sum_{k=1}^{n-1} \beta_k(s) \ket{v_k(s)}) \ket{\bar{0}0^{\tau}00}\| \tag*{} \\
		&=\| \Pi_{\bar{0}0} ( W(s)_{\tau} \x I)(\sum_{k=1}^{n-1} \beta_k(s) \ket{v_k(s)}) \ket{\bar{0}0^{\tau}00}\|\tag*{}.
	\end{align}

	Now we consider the first item in \cref{eq:sss}. Since $\ket{\pi} = \sqrt{1-\pi_g}\ket{\bar{\pi}}+\sqrt{\pi_g}\ket{g}$ and $\left \langle \bar{\pi} | g\right \rangle = 0$, combine with \cref{eq:v0} we have
	$$\left \langle \pi | v_0(s)\right \rangle  = \langle{\bar{\pi}}|{v_0(s)}\rangle\sqrt{1-\pi_g} + 	\langle{g}|{v_0(s)}\rangle\sqrt{\pi_g} = \sqrt{\frac{(1-s)(1-\pi_g)}{1-s(1-\pi_g)}}\sqrt{1-\pi_g} + \sqrt{\frac{\pi_g}{1-s(1-\pi_g)}}\sqrt{\pi_g},$$
	$$\beta_0(s)=\left \langle v_0(s) | g\right \rangle  = \sqrt{\frac{\pi_g}{1-s(1-\pi_g)}}.$$
	Let	$s = 1 - \frac{\pi_g}{1-\pi_g}$, the above equation will become
	$\left \langle \pi | v_0(s)\right \rangle \ge(\sqrt{1-\pi_g}+\sqrt{\pi_g})/\sqrt{2}\ge1/\sqrt{2},\beta_0(s)\ge1/\sqrt{2}$.
	Thus the first item of \cref{eq:sss} is $| \beta_0 \left \langle \pi | v_0(s)\right \rangle  |\ge 1/2$.	Now consider the second item in \cref{eq:sss}, which can be expressed as 
	\begin{align}
		&\| \Pi_{\bar{0}0} ( W(s)_{\tau} \x I)(\sum_{k=1}^{n-1} \beta_k(s) \ket{v_k(s)}) \ket{\bar{0}0^{\tau}00}\|  \tag*{}  \\
		&\le \| \sum_{k=1}^{n-1} \beta_k(s)(D^{t}\ket{v_k(s)})\ket{\bar{0}0^{\tau}00}\|  +  \| \Pi_{\bar{0}0}( W(s)_{\tau} \x I)(\sum_{k=1}^{n-1} \beta_k(s)\ket{v_k(s)})\ket{\bar{0}0^{\tau}00} - (\sum_{k=1}^{n-1} \beta_k(s)(D^{t}\ket{v_k(s)}))\ket{\bar{0}0^{\tau}00} \|.
		\label{eq:as}
	\end{align}
	Let $t = \log(\frac{4}{\varepsilon})\frac{2}{\delta(s)}$ and $\varepsilon_{1} = \frac{\varepsilon}{2}$, combine with \cref{eq:close} and \cref{eq:ccos}, the above equation can be expressed as 
	$$2 | \sum_{k=1}^{n-1} \cos^{t}(\varphi_k)\beta_k(s)|  +  \frac{\varepsilon}{2}
	\le \frac{\varepsilon}{2}| \sum_{k=1}^{n-1}\beta_k(s)|  + \frac{\varepsilon}{2}	\le \varepsilon.$$
	Then the success probability of \cref{alg:qis} is a constant approaching to $1/4$.
	
	\textbf{The complexity of algorithm with known $\pi_g$}
	
	Since $W(s)$ corresponds to the graph with single-marked vertex $g$ and $\Theta (\sqrt{\Heg}) = \Theta(1/\sqrt{\delta(s)})$\footnote{See Equation(86) and Equation(87) \cite{PhysRevA.102.022423} to find $\Theta(1/\delta(s))=  \Theta (\Heg)$ when  $|M|=1$.}, $W(s)$ is applied $\Theta(\Gamma) = \Theta (\log(1/\varepsilon)\sqrt{\Heg})$ times in total.
	From \cref{thm:2}, the number of ancilla qubits required in \cref{alg:qis} is $\tau = \lceil \log\Gamma\rceil= \Theta(  \log(\frac{1}{\sqrt{HT}}) + \log\log(\frac{1}{\varepsilon}) )$.
	\begin{remark}
		\label{rem:lazy}
		The effect of introducing lazy walk on complexity is only related to gap in \cref{thm:qsampling with known pm}. From the definition of lazy walk, the eigenvalue of lazy walk can be expressed as $\lambda^{lazy}_j = (1+\lambda_j)/2$ for $j = 0,1,\cdots,n-1$, and then gap becomes $\delta^{lazy} = \lambda^{lazy}_0 - \lambda^{lazy}_1=(\lambda_0 - \lambda_1)/2=\delta/2$, which only affects complexity by adding a constant factor $2$.
	\end{remark}	    
	
	Since qsampling requires accuracy as the form of $1 - \varepsilon_0$, a constant overlap with stationary state $\ket{\pi}$ is not sufficient. Here we take the method of amplitude amplification %\footnote{To avoid the souffl$\acute{e}$ problem, {\color{red}{the authors introduce the method of fixed-point amplitude amplification with quadratic speedup over classical case instead}} of amplitude amplification. The detailed construction can be found in \cite{Yoder2014FixedpointQS}, which differs from amplitude amplification in changing the phase of each reflection on the initial state and target state.} 
	to enlarge the success probability. The reflection on $\ket{\pi}$ is prepared in \cref{thm:2}, and the reflection on the current state $\ket{\psi_m}$ can be prepared as 
	$$R_m=I-2\ket{\psi_m}\bra{\psi_m}=U_{main}(I-2\ket{g}\bra{g})U_{main}^{\dagger}=U_{main}R_gU_{main}^{\dagger},$$ which is shown in \cref{fig:a00}. Combine them and we will prepare a state $\varepsilon_0$-close to $\ket{\pi}$, and the total complexity of qsampling from the stationary state is $\Theta(\sqrt{\Heg}\log(1/\varepsilon_0))$.
	
	\subsection{The case with unknown $\pi_g$} \label{sect:unknown pm}
	\hspace{1.2em}
	It seems that we have solved the qsampling problem, however, there still exists an unsolved case.
	Above algorithm requires knowing the value of $\pi_g$ corresponding to the initial state $\ket{g}$. Since $\ket{\pi}$ is the target state and $\pi_g$ is component of $\ket{\pi}$, $\pi_g$ is not always known without other information. A new algorithm to determine the value of $\pi_g$ is constructed as below.
	A common method for determining $\pi_g$ is binary search. 
	By narrowing the problem down to a smaller interval, we optimize the algorithm and reduce the complexity of the algorithm.
	For this, we need the following results from Krovi et al. \cite{Krovi2015QuantumWC}, which shows that when $\pi_g$ and its approximate value $\pi^*$ are close enough, $\ket{\pi}$ can be prepared by $s$ determined from $\pi^*$, and success probability (the first term of \cref{eq:sss}) is as high as $1/9$.
	%	We consider the case with an approximate value of $\pi_g$ firstly.
	\begin{lemma}[Proposition 5 in~\cite{Krovi2015QuantumWC}]
		For arbitrary $\pi_g \in (0,\frac{1}{2}) $, assume there is an approximate value of $\pi_g$ as $\pi^*$,  if
		\begin{equation}	\label{eq:approximate p}
			|\pi^* - \pi_g| \leq \pi_g/3
		\end{equation}
		then both $\left \langle v_0(s^*)  | \bar{\pi}\right \rangle\left \langle v_0(s^*)  | g\right \rangle \geq \frac{1}{3}$ and $\left \langle v_0(s^*) | g\right \rangle ^2  \geq \frac{1}{3}$ holds, where $s^*= 1 - \frac{\pi^*}{1-\pi^*}$.
		\label{lem:appro}\end{lemma}
	Next we consider how to prepare the stationary state $\ket{\pi}$ with unknown $\pi_g$ as below.
	\begin{theorem}
		\label{thm:qsampling with bound on p}
		For any ergodic reversible Markov chain P in $G(V,E)$, $\ket{\pi}$ can be prepared by		\cref{alg:unknown} with constant success probability, and the corresponding complexity is $\Theta( \log(\pi_{min}^{-1}/n)\sqrt{\Heg}\log\frac{1}{\varepsilon})$, where $\varepsilon$ is error. \end{theorem}
	
	\begin{figure}[H]
		\centering 
		\includegraphics[width=1.3\textwidth,trim=42 0 0 0,clip]{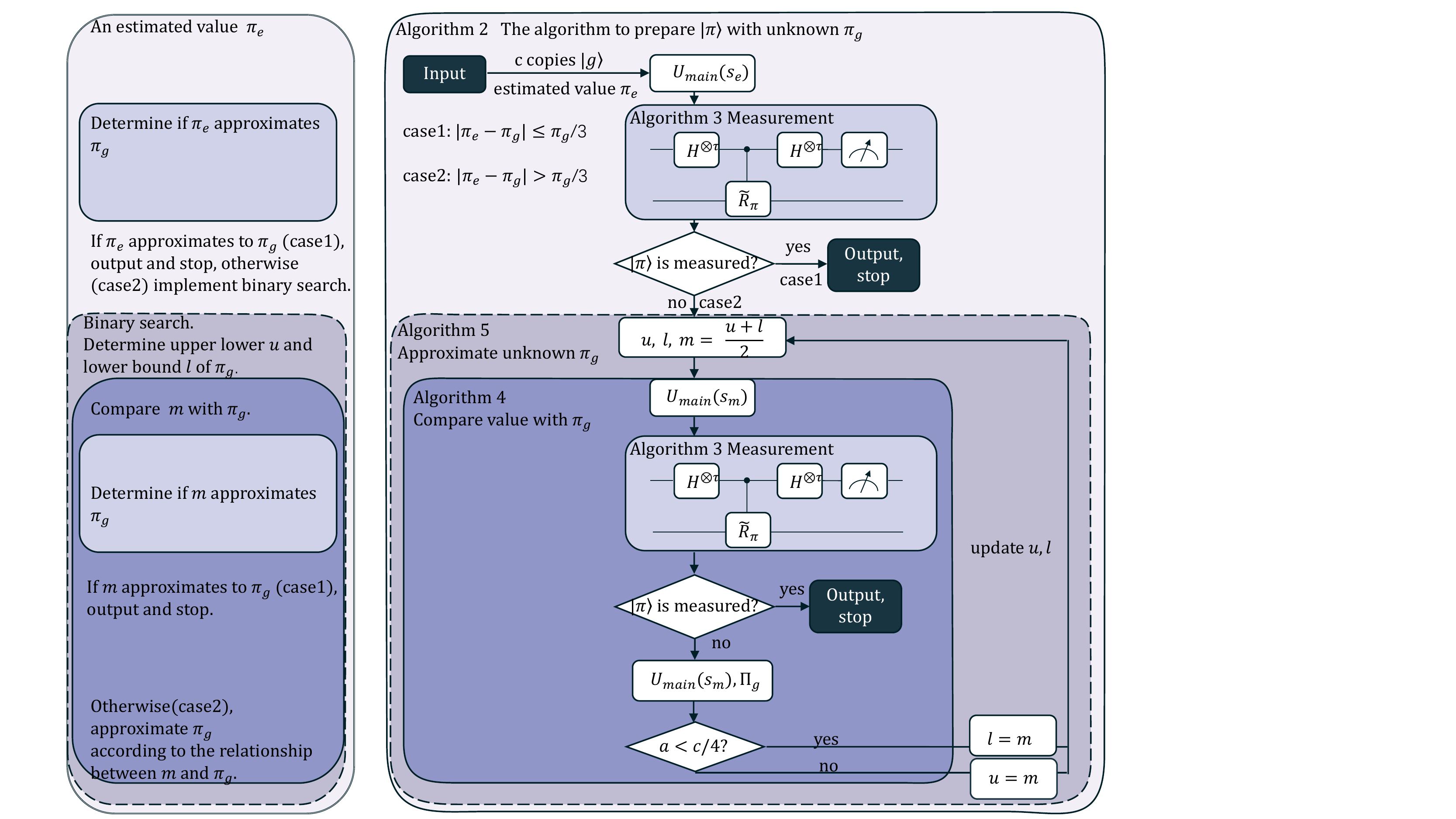}
		\caption{Main algorithm framework to prepare $\ket{\pi}$ with unknown $\pi_g$, where the left is goal and the right is algorithm framework.}
		\label{fig:a4}
	\end{figure}
	
	\begin{algorithm}[H]
		\caption{The algorithm to prepare $\ket{\pi}$ with unknown $\pi_g$}
		\label{alg:unknown}
		\textbf{Input}: $\varepsilon$, the expected error; $L=\Theta(\log(\pi_{min}^{-1}/n))$, the maximum times to apply binary search.\\
		\textbf{Output}: a state $\varepsilon$-close to $\ket{\pi}$; $\pi^*$, the estimated value of $\pi_g$.
		
		\begin{algorithmic}[1]
			\STATE Choose vertex $g$ in $V$ randomly, and set $\pi_e = 1/n$.
			\STATE Set $c$ copies of initial state as $\ket{g} \ket{\bar{0}0^{\tau}00}$ on $\Reg_1\Reg_2\Reg_3\Reg_4\Reg_5$, where the fourth registers are the same as \cref{thm:2} and the fifth register $\Reg_5$ is a single-qubit register as $\spn\{ \ket{0}, \ket{1}\}$.
			\FOR {$k = 1,2,\dots,c$}\label{step:00}
			\STATE Apply $U_{main}(s_e)$ to the $k$th initial state, where $M =\{g\}$, $s_e = 1-\frac{\pi_e}{1-\pi_e}$, $\Gamma = \Theta(\sqrt{\Heg}\log\frac{1}{\varepsilon})$.
			\STATE Perform \cref{alg:check} to measure whether the state in $\Reg_1$ is $\ket{\pi}$. \label{step:02}
			\IF {$\ket{\pi}$ is measured}
			\RETURN $\ket{\pi}$ and $\pi_e$, stop.
			\ENDIF
			\ENDFOR\label{step:01}
			\STATE Call \cref{alg:search}$(g, 0, 100/n, L)$. 
		\end{algorithmic}
	\end{algorithm}
	
	\textbf{The description of algorithm with unknown $\pi_g$(\cref{alg:unknown})}
	
	The goal of \cref{alg:unknown} is to estimate the value of $\pi_g$. We assume the estimated value of $\pi_g$ as $\pi_e = 1/n$ and  use \cref{alg:unknown} to prepare $\ket{\pi}$ with unknown $\pi_g$ as follows, and the corresponding construction can be seen in	\cref{fig:a4}. It can be divided into two cases.
	
	Case 1: $\pi_e$ approximates $\pi_g$, $2\pi_g/3 \le \pi_e \le  4\pi_g/3$.
	
	Case 2: $\pi_e$ does not approximate $\pi_g$, $\pi_e <  2\pi_g/3$ or $\pi_e >  4\pi_g/3$.
	
	The process to determine if $\pi_e$ approximates to $\pi_g$ is in \Step{00}-\Step{01} of \cref{alg:unknown}, and here we introduce the measurement in \Step{02} firstly.
	
	\textbf{Measure whether the state in $\Reg_1$ is $\ket{\pi}$(\Step{02} in \cref{alg:unknown})}
	
	To measure whether the current state is $\ket{\pi}$, we need the projection operator $\Pi_{\pi} = \ket{\pi}\bra{\pi}$. Use the approximate reflection $\tilde{R}_{\pi}$ on $\ket{\pi}$ in \cref{thm:2} to construct $\Pi_{\pi}$ and the  process is shown in \cref{alg:check}.
	
	\begin{algorithm}[H]	
		\caption{Algorithm to measure whether the state in $\Reg_1$ is $\ket{\pi}$}
		\label{alg:check}
		\textbf{Input}: The state $\ket{\psi_c}\ket{\cdots}$ in $\Reg_1\Reg_2\Reg_3\Reg_4\Reg_5$, where $\ket{\psi_c}$ will be measured to be $\ket{\pi}$ or not.\\
		\textbf{Output}: ``$\ket{\pi}$ is measured" or ``fail".
		
		\begin{algorithmic}[1]
			\STATE Append $\Reg_{check}$ with initial state $\ket{0^{\tau+1}}\ket{0}$  as ancilla register.
			\STATE Apply Hadamard gate on $\Reg_{check}$.\label{step:2}
			\STATE Apply controlled-$\tilde{R}_{\pi}$ on $\Reg_1\Reg_2$ when $\Reg_{check}$ is $\ket{0}$.
			\STATE Apply Hadamard gate on $\Reg_{check}$ and measure $\Reg_{check}$. \label{step:4}
			\IF {$\Reg_{check} = 0$}
			\RETURN ``$\ket{\pi}$ is measured".
			\ELSE
			\RETURN ``fail".
			\ENDIF
		\end{algorithmic}
	\end{algorithm}	
	A short description of \cref{alg:check} is given as below.
	Let $\ket{\psi_{c}}\ket{0}$ denote the current state on $\Reg_1\Reg_{check}$, the controlled-$\tilde{R}_{\pi}$ in the first register can be seen as an approximate controlled reflection on $\ket{\pi}$, and then \Step{2} - \Step{4} can be approximately expressed as
	\begin{align*}& ( I\x H )((2\ket{\pi}\bra{\pi}-I )\x\ket{0}\bra{0} + I\x \ket{1}\bra{1} )( I\x H)\ket{\psi_{c}}\ket{0}\\
		=& \langle \pi  | \psi_c \rangle  \ket{\pi}\ket{0}  +\langle \pi^{\perp}  | \psi_c \rangle\ket{\pi^{\perp} } \ket{1}.\end{align*}
	Thus, if $0$ is measured in $\Reg_{check}$, the current state in $\Reg_1$ is $\ket{\pi}$, and the probability of $0$ in $\Reg_{check}$ is $| \langle \pi  | \psi_c \rangle |^2$.
	Then we can determine whether the state in $\Reg_1$ is $\ket{\pi}$ by applying \cref{alg:check}.
	
	\textbf{Determine if $\pi_e$ approximates to $\pi_g$(\Step{00} - \Step{01} in \cref{alg:unknown})}
	
	If $\pi_g$ and $\pi_e$ satisfy case 1, $\ket{\pi}$ will be achieved by $U_{main}$ with at least $1/9$ success probability by \cref{lem:appro}. 
	For $c$ initial states acted by $U_{main}$ with $s_e=1-\frac{\pi_e}{1-\pi_e}$, measure the first register of $c$ states by \cref{alg:check} , at least one state should be measured as $\ket{\pi}$ with probability more than $1-(8/9)^c$, which is nearly 1 when $c = 100$. 
	Thus if no $\ket{\pi}$ in $c$ states is measured, we think $\pi_e$ does not approximate $\pi_g$.
	
	We distinguish case 1 from case 2 by whether $\ket{\pi}$ is measured. If $\pi_e$ approximates $\pi_g$ (case1), \cref{alg:unknown} output and stop; otherwise, $\pi_e$ does not approximate $\pi_g$ (case2) and we implement binary search as \cref{alg:search} to approximate $\pi_g$.

	In binary search, we determine the upper bound $u$ and lower bound $l$ of $\pi_g$ firstly. 
	Here we assume $\pi_g \in [l, u] = [0, C/n]$, and let the intermediate value $m = (u + l)/2$.
	The next step is to compare the intermediate value $m$ with $\pi_g$. In \cref{alg:compare} we give the process to compare any value $x$ with $\pi_g$, which is the necessary premise to implement binary search.
	
	\begin{algorithm}[H]
		\caption{Compare any value $x$ with $\pi_g$}
		\label{alg:compare}
		\textbf{Input}: $g$, the initial state vertex; $x$, the value to be compared with $\pi_g$.\\
		\textbf{Output}: “$x< 2\pi_g/3$” or  “$x > 4\pi_g /3$” or “$|x - \pi_g| \leq \pi_g/3$” with $\ket{\pi}$.
		
		\begin{algorithmic}[1]
			\STATE Set $c$ copies of initial state as $\ket{g} \ket{\bar{0}0^{\tau}00}$ on $\Reg_1\Reg_2\Reg_3\Reg_4\Reg_5$, where the fourth registers are the same as \cref{thm:2} and the fifth register $\Reg_5$ is a single-qubit register as $\spn\{ \ket{0}, \ket{1}\}$.
			\FOR {$k = 1,2,\dots,c$}\label{step:03}
			\STATE Apply $U_{main}(s_x)$ to the $k$th initial state, where $M =\{g\}$, $s_x = 1-\frac{x}{1-x}$, $\Gamma = \Theta(\sqrt{\Heg}\log\frac{1}{\varepsilon})$.
			\STATE Perform \cref{alg:check} to measure whether the state in $\Reg_1$ is $\ket{\pi}$. 
			\IF {$\ket{\pi}$ is measured}
			\RETURN “$|x - \pi_g| \leq \pi_g/3$” and $\ket{\pi}$, stop.
			\ENDIF
			\ENDFOR\label{step:04}
			\STATE Rebuild $c$ copies of initial state as $\ket{g} \ket{\bar{0}0^{\tau}00}$ on $\Reg_1\Reg_2\Reg_3\Reg_4\Reg_5$, $a=0$.\label{step:05}
			\FOR {$k = 1,2,\dots,c$}
			\STATE Apply $U_{main}(s_x)$ to the $k$th initial state, where $M =\{g\}$, $s_x = 1-\frac{x}{1-x}$, $\Gamma = \Theta(\sqrt{\Heg}\log\frac{1}{\varepsilon})$.\label{step:gg}
			\STATE Measure with $\Pi_{g}=\ket{g}\bra{g}$. \label{step:ggg}
			\IF {$\ket{g}$ is measured}
			\STATE a = a + 1.
			\ENDIF
			\ENDFOR
			\IF {$a \ge c/4$}
			\RETURN “$x < 2\pi_g /3$”.
			\ELSE
			\RETURN “$x > 4\pi_g /3$”.
			\ENDIF\label{step:06}
		\end{algorithmic}
	\end{algorithm}
	
	If $x$ and $\pi_g$ satisfy \Step{03} and \Step{04},  at least one $\ket{\pi}$ is measured in $c$ copies state with probability nearly 1 by \cref{lem:appro} and the algorithm stops; otherwise, no $\ket{\pi}$ being measured means that the current estimated value is not close enough to $\pi_g$, and the next problem is to distinguish $x < 2\pi_g/3$ from  $x > 4\pi_g /3$. 
	
	\textbf{Distinguish $x < 2\pi_g/3$ from  $x > 4\pi_g /3$(\Step{05} - \Step{06} in \cref{alg:compare}) }
	
	We apply $U_{main}(s_x)$ and measure the state in $\Reg_1$ with $\Pi_{g}$ firstly.
	Let $p^g$ denote the probability that $g$ is measured, from the proof in \cref{thm:qsampling with known pm} we have	
	\begin{align*}	\sqrt{p^g}	\label{eq:pg} 		
		&= \| (\ket{g}\bra{g}_1 U_{main} (\ket{g}\ket{\bar{0}0^{\tau}00})\|\tag*{ }\\
		&\ge \| (\bra{g}\bra{\bar{0}0^{\tau}}\x I\x I) U_{main} (\ket{g}\ket{\bar{0}0^{\tau}00})\|\tag*{ }\\
		&\ge \|\left \langle g | v_0(s_x)\right \rangle  \left \langle v_0(s_x) | g\right \rangle \| - \| \Pi_{\bar{0}0} W(s_x)_{\tau} (\sum_{k=1}^{n-1} \beta_k(s_x) \ket{v_k(s_x)}) \ket{\bar{0}0^{\tau}}\|  \\
		&\ge  \|\left \langle g | v_0(s_x)\right \rangle  \|^2	- \varepsilon = \beta_0^2(s_x)	- \varepsilon,	
	\end{align*}
	where $\varepsilon$ is arbitrarily small. Then the ratio of $a$ (the time $g$ is measured) to $c$ (total number of states) will be close to $p^g \approx  \beta^4_0(s_x)$.
	
	Since $s_x= 1 - \frac{x}{1-x}$ varies with the value $x$, from \cref{eq:v0} and \cref{eq:g}, if $x \ge 4/3\pi_g$ we have:
	\begin{align*}
		p^g_{x \ge 4/3\pi_g} \approx \beta^4_0(s_x)	
		= ( \frac{\pi_g(1-x)}{\pi_g+x-2\pi_g x})^2
		\le ( \frac{\pi_g(1-4\pi_g/3)}{\pi_g+4\pi_g /3-2\pi_g  \cdot4\pi_g/3})^2
		%= ( \frac{1}{2}-\frac{1}{2}\cdot\frac{1}{7-8\pi_g})^2
		%\le  (\frac{1}{2}+\frac{1}{2}\cdot\frac{1}{7})^2
		\le  \frac{9}{49}.
	\end{align*}
	Similarly, if $x \le 2/3\pi_g$ we have
	$p^g_{x \le 2/3\pi_g} \approx \beta^4_0(s_e)\ge  \frac{9}{25}.$

	Let a number that approximates the mean of $\frac{9}{49}$ and $\frac{9}{25}$ be the dividing point between $x < 2\pi_g/3$ and $x > 4\pi_g /3$. Here we choose $1/4$. Then $x < 2\pi_g/3$ is chosen if more than $c/4$ copies measured as $\ket{g}$; otherwise, the result is chosen as $x > 4\pi_g/3$.
	%	Let $1/4$, a number that is about the mean of $\frac{9}{49}$ and $\frac{9}{25}$, be the dividing point between $x < 2\pi_g/3$ and $x > 4\pi_g/3$, which means if there are more than $c/4$ copies measured as $\ket{g}$ in $c$ copies state applied by  $U_{main}(s_x)$, we think $x < 2\pi_g/3$; otherwise, the result is $x > 4\pi_g/3$.
	
	We now give a lower bound on the success probability of the judgment that distinguishes $x < 2\pi_g/3$ from  $x > 4\pi_g /3$. Define independent and identically distributed variables $X_{1}, X_{2}, \cdots, X_{r}$ that satisfies
	$P(X_{i} = 1) = p, P(X_{i} = 0) = 1-p.$
	
	If $x > 4\pi_g/3$, the error only occurs when $a<c/4$. Let $X_{i}=1$ denote the case that the $i$th measurement result is $\ket{g}$, then $$p \coloneqq Pr(\ket{g} \ is \ measured \ in \ ith \ measurement) = p^g \le \beta^4_0(s_e)  \le \frac{9}{49}.$$ 
	By Chebyshev's inequality, the error probability $p_{error}$ can be expressed as
	$$p_{error} 
	= P(|\bar{p}-p|\ge e) 
	\le\frac{p(1-p)}{le^2} 
	\le\frac{1}{4},$$
	where 	$e =|9/49-1/4|$, $\bar{p} = \frac{1}{r} {\textstyle \sum_{i=1}^{r}X_{i}}$, and $r=100$.
	
	Similarly, the error only occurs when $a\ge c/4$ if $x < 2\pi_g/3$, and the time that $\ket{g}$ is not measured is $c-a \le 3c/4$. 
	By Chebyshev's inequality, the error probability $p_{error}$ can be expressed as
	$$p_{error} 
	= P(|\bar{p}-p|\ge e) 
	\le\frac{p(1-p)}{le^2} 
	\le\frac{1}{4},$$
	where 	$e =|3/4-16/25|$, $\bar{p} = \frac{1}{r} {\textstyle \sum_{i=1}^{r}X_{i}}$, and $r=100$.
	
	After the above discussion, we see the probability that \cref{alg:compare} successfully judges the relationship between $x$ and $\pi_g$ is greater than 3/4 whether $x \le 2/3\pi_g$ or $x \ge 4/3\pi_g$.
	And then we use the above comparison subroutine to binary search.
	
	\textbf{Binary search(\cref{alg:search})}
	
	In binary search of \cref{alg:search}, \cref{alg:compare} is called to compare the value of  each lower bound $l(i)$, upper bound $u(i)$, $m(i)$ with $\pi_g$.
	If any value approximates $\pi_g$, the algorithm outputs $\ket{\pi}$ and stops; otherwise, the relationship between the value and $\pi_g$ is given.
	If the results of lower bound $l(i)$ and upper bound $u(i)$ contradict, backtrack to the previous values, else compare $m[i]$ with $\pi_g$ and set $m[i]$ as either $u[i+1]$ or $l[i+1]$ according to the relationship between $m[i]$ and $\pi_g$.
	
	Besides, we define a maximum number of times as $L$ in binary search. When $\ket{\pi}$ is still not prepared after $L$ times, the algorithm stops and outputs ''fail''.
	
	The above is the description of \cref{alg:unknown}, and we compute the success probability of \cref{alg:unknown} as below.
	
	\begin{algorithm}[H]
		\caption{Approximate unknown $\pi_g$ by binary search}
		\label{alg:search}
		\textbf{Input}: $g$, the initial state vertex; $l$, the lower bound of $\pi_g$; $u$, the upper bound of $\pi_g$; L, the maximum search times. \\
		\textbf{Output}: $\pi^*$, the approximate value of $\pi_g$ or "fail".
		\begin{algorithmic}[1]
			\STATE Set $i = 0$, $l[0] =l$, $u[0] = u$, $m[0]=  \lceil (l+ u )/2 \rceil$.
			\WHILE{$i\le L$ and $i\ge 0$}
			\STATE Call \cref{alg:compare}$(g, l[i])$, \cref{alg:compare}$(g, u[i])$.\label{step:e}
			\IF {return “$|l[i] - \pi_g| \leq \pi_g/3$” with $\ket{\pi}$, or “$|u[i] - \pi_g| \leq \pi_g/3$” with $\ket{\pi}$}
			\STATE Output the current value as $\pi^*$, stop.
			\ELSIF {return “$l[i] \ge 4\pi_g /3$” or “$u[i] \le 2\pi_g /3$”}
			\STATE Call \cref{alg:search}$(g, l[i-1], u[i-1])$.
			\ELSE 
			\STATE Call \cref{alg:compare}$(g, m[i])$.	
			\IF {return “$|m[i] - \pi_g| \leq \pi_g/3$” with $m[i]$}
			\STATE Output the value as $\pi^*$, stop.
			\ELSIF {return “$m[i] \le 2\pi_g /3$”}
			\STATE Call \cref{alg:search}$(g, m[i], u[i], L)$, $i = i + 1$.
			\ELSE
			\STATE Call \cref{alg:search}$(g, l[i], m[i], L)$, $i = i + 1$.
			\ENDIF
			\ENDIF
			\ENDWHILE
			\RETURN ``fail".
		\end{algorithmic}
	\end{algorithm}	
	
	\textbf{The success probability of algorithm with unknown $\pi_g$}
	
	Since the probability of success can be expressed as 1 minus the probability of failure, we compute the probability of failure as below.
	
	That algorithm fails means $\ket{\pi}$ is not measured in \cref{alg:unknown}.  Let $S_k$ denote that ``$\ket{\pi}$ is not measured in step.k", then the corresponding probability can be calculated as follows:
	\begin{align*}
		&	Pr(\ket{\pi} \ is \ not \ measured)	=Pr(S_5,S_{10})\\
		=&Pr(S_5,S_{10},\pi_g >  C/n)+Pr(S_5,S_{10},\pi_g \le C/n)\\
		\le&Pr(\pi_g > C/n)+ Pr(S_5, S_{10}, |\pi_g-1/n|\le\pi_g/3,\pi_g \le C/n)+Pr(S_5, S_{10}, |\pi_g-1/n|>\pi_g/3,\pi_g \le C/n)\\
		\le&Pr(\pi_g > C/n)+Pr(S_5, |\pi_g-1/n|\le\pi_g/3)+Pr(S_{10},\pi_g \le C/n).
	\end{align*}
	The upper bounds of these terms are given below. 
	
	\textbf{Case1: $\pi_g > C/n$.}
	
	First consider the value of $Pr(\pi_g > C/n)$.	From the property of stationary distribution we have	\begin{align*}	
		1&=\sum_{y\in V}\pi_y
		=\sum_{\pi_y > C/n, y\in V}\pi_x+\sum_{\pi_y \le C/n, y\in V}\pi_y
		\ge \sum_{\pi_y > C/n, y\in V}\frac{C}{n}+\sum_{\pi_y \le C/n, y\in V}\pi_y
		\ge |\{y|\pi_y > \frac{C}{n}, y\in V\}|\frac{C}{n}.
	\end{align*}
	Equivalently, $|\{y|\pi_x > \frac{C}{n}, y\in V\}|\le \frac{n}{C}$, so  \begin{equation}
		Pr(\pi_g > C/n) = \frac{|\{y|\pi_y > C/n, y\in V\}|}{|\{y| y\in V\}|} \le \frac{n/C}{n} = 1/C.
	\end{equation}
	Since $C$ can be chosen as a large constant, then the probability that a vertex $g$ is randomly chosen from $V$ with $\pi_g = \Omega (1/n)$ is at least $1-1/C$. 
	
	\textbf{Case2: $S_5, |\pi_g-1/n|\le\pi_g/3$.}
	
	Second bound the value of $Pr(S_5, |\pi_g-1/n|\le\pi_g/3)$.
	If $|\pi_g-1/n|\le\pi_g/3$, $\ket{\pi}$ will be achieved by $U_{main}$ with at least $1/9$ success probability by \cref{lem:appro}. This means for $c$ initial states acted by $U_{main}$, measure the first register of $c$ states and the probability that $\ket{\pi}$ is not measured in \Step{02} is less than $(8/9)^c=(8/9)^{100}$.
	
	\textbf{Case3: $S_{10},\pi_g \le C/n$.}
	
	Now we bound the value of $Pr(S_{10},\pi_g \le C/n)$. The corresponding number of digits in the possible interval is $N = \frac{C/n}{\pi_{min}}=\Theta(\frac{1}{n\cdot\pi_{min}})$ since $\pi_g\in[0, C/n]$. Define $L\coloneqq\Theta(\log(\pi_{min}^{-1}/n))$.
	From \cite{Feige1994ComputingWN}, for every $Q < 1/2$, the binary search algorithm computes the correct value of $\pi_g$ with probability at least $1-Q$ in $O(log(N/ Q))$ steps, where $N$ is the number of digits in the possible interval.
	
	Here we have $Q\approx1/4$. Therefore, the probability that $\pi_g$ is found in $[0, C/n]$ with number of steps more than $L$ is less than $Q$, that is $Pr(S_{10},\pi_g \le C/n)\le Q$.
	
	The bound of above cases have been given, set $C=100$ and we have $$Pr(\ket{\pi} \ is \ not \ measured)	 \le	1/C+(1-1/3)^c+Q\le 1/100+(1-1/3)^{100}+1/4<3/10.$$
	Then  the success probability is at least $7/10$.
	
	Assume it needs $t$ times of calling \cref{alg:unknown} and we have 
	\begin{align*}
		t& \le \sum_{i=1}^{\infty} i\times\frac{7}{10}\times{\frac{3}{10}}^{i-1} = \frac{7}{10}[\sum_{i=1}^{\infty} x^i]'|_{x=3/10} 
		= \frac{7}{10}[\frac{x}{1-x}]'|_{x=3/10} 
		= \frac{7}{10}[\frac{1}{(1-x)^2}]|_{x=3/10} 
		= \frac{10}{7}<2
	\end{align*}
	Namely, we just run $1$ time of \cref{alg:unknown} to get a higher probability.

	\textbf{The complexity of algorithm with unknown $\pi_g$}
	
	In \cref{alg:unknown}, the main operators used are $U_{main}$ and $\tilde{R}_{\pi}$.
	
	Since each $U_{main}$ calls  $\Theta(\log(1/\varepsilon)\sqrt{\Heg})$ times walk operator $W$ and $U_{main}$ is invoked $\Theta(\log(\pi_{min}^{-1}/n))$ times in total, the number of times that $W$ is called can be expressed as $\Theta(\log(\pi_{min}^{-1}/n)\log(1/\varepsilon)\sqrt{\Heg})$.
	
	What's more, we consider $\tilde{R}_{\pi}$.
	From \cref{thm:2}, each $\tilde{R}_{\pi}$ calls  $\Theta(\log(1/\varepsilon)\sqrt{\delta})$ times walk operator $W$. Then the total complexity related to $\tilde{R}_{\pi}$ is $\Theta(\log(\pi^{-1}_{min}/n)cost(R_{\pi}))=\Theta(\log(\pi^{-1}_{min}/n)(\log(1/\varepsilon)\sqrt{\delta})))$.
	
	In summary,  the total complexity in \cref{alg:unknown} is 
	$$\Theta(\log(\pi_{min}^{-1}/n)\log(1/\varepsilon)\sqrt{\Heg})+\Theta(\log(\pi_{min}^{-1}/n)\log(1/\varepsilon)\sqrt{\delta})).$$
	Since from the relationship between classical mixing time and hitting time we always have $\sqrt{\delta}\le\sqrt{\Heg}$, then the total complexity is $$\Theta(\log(\pi_{min}^{-1}/n)\log(1/\varepsilon)\sqrt{\Heg}+\log(\pi^{-1}_{min}/n)\log(1/\varepsilon)\sqrt{\delta}))=\Theta\left(\log(\pi_{min}^{-1}/n)\log(\varepsilon^{-1})\sqrt{\Heg}\right).$$
	
	\begin{remark}
		Erd\"os-R\'enyi random graph $G(n, p)$ is obtained from the complete graph of $n$ vertices by keeping each edge with probability $p$. Each edge is independent of all other edges and has the same probability. Let $D_x$ represent the number of vertices connected to $x$, then the probability from vertex $x$ to each connected vertex is $1/D_x$. For a pair of connected vertices $x$ and $y$, we have $p_{xy} = 1/D_x$, $p_{yx} = 1/D_y$.
		
		Let $c$ denote a constant and $p=c/n$, and $G(n, p)$ can be seen as sparse graphs. For arbitrary vertex $x$, the expectation of  $D_x$ is $c$, which means $D_x=c_1$, $D_y=c_2$,  $c_1$ and $c_2$ are constants. For any pair of connected vertices $x$ and $y$, due to the properties of reversible Markov chains, $\pi_x= \pi_y D_x/D_y=\pi_y c_1/c_2= \pi_y c_0$, where $c_0\coloneqq c_1/c_2$ is a constant as well.  Since $1= \sum_{x\in V}\pi_x=\sum_{x \in V}\pi_{min}D_{min}/D_x=\pi_{min}(\sum_{x \in V}D_{min}/D_x)= \pi_{min}(\sum_{x \in V}c_x)$, where each $c_x$ is a constant. Thus $\sum_{x \in V}c_x=\Theta(n)$, $\pi_{min}=\Theta(1/n)$,  the complexity of algorithm to prepare $\ket{\pi}$ in sparse graphs is $\Theta(\sqrt{\Heg}\log(1/\varepsilon))$.
	\end{remark}
	\begin{remark}
		The above algorithms are suitable to the case with small $\pi_g$, $i.e.$ $\pi_g < \Theta(1)$. For any reversible Markov chains from the initial state $\ket{g}_{g \in V}$ with $\pi_g \ge c$, where $c$ is a constant, it is faster to use amplitude amplification directly. The corresponding complexity is constant times of calling $\tilde{R}_{\pi}$ and $R_g$, that is $\Theta(\sqrt{\delta}\log(\varepsilon_{1}^{-1}))$, and achieves a quadratic speedup compared with classical mixing time.
	\end{remark}

	\section{Discussion} \label{sect:Discussion}
	\hspace{1.2em}
	When referring to qsampling algorithm, most existing results consider the Markov chains corresponding to special graphs, such as lattice \cite{Richter2007QuantumSO}, circle, and some special stationary distributions \cite{Dunjko2015}. 
	
	We list the state of the art firstly. Consider discrete-time quantum algorithm, by reversing search algorithm \cite{Krovi2015QuantumWC, Li2020BoundsOQ}, the corresponding complexity is $\Theta(\sqrt{\Heg}\frac{1}{\varepsilon})$. In continue-time quantum algorithm, by Hamiltonian evolution with von Neumann measurements, stationary state is prepared from $\ket{g}$ ($g \in V$) with known $\pi_g$, and the corresponding complexity is $\Theta(\sqrt{\Heg}\log\frac{1}{\varepsilon})$ \cite{PhysRevA.102.022423}. 
	However, since $\ket{\pi}$ is the state we want and $\pi_g$ is part of $\ket{\pi}$, the process of how to get the value of $\pi_g$ should not be ignored.
%	We consider the non-regular graphs (the case with unknown $\pi_g$) and speedup the qsampling of non-regular graphs in both discrete-time case and continue-time case. Besides, the number of ancilla qubits required is reduced.
	In this work, we propose a new qsampling algorithm by combining quantum fast-forwarding algorithm with interpolated walk for all reversible Markov chains, which does not restrict the form of Markov chain or limit  stationary distribution. Our qsampling algorithm starts from the initial state corresponding to vertex randomly selected and display a speedup compared with the previous quantum algorithms on non-regular graphs. In some common graphs, especially for sparse graphs that are difficult to quickly reach the stationary distribution, we achieve a complete quadratic speedup compared with classical mixing time(without logarithmic factor), and our algorithm improves the sampling preparation in both quantum case and classical case.
	
	One of the most useful applications of our algorithm is to prepare stationary state in sparse graphs. Since the sparser the graph, the larger its diameter, and the more difficult it is to mix quickly. Besides, sparse graphs are not as regular as circles and grids. Arbitrarily choose a vertex $g$ in the sparse graph, the corresponding $\pi_g$ is unknown.
	Consider Erd\"os-R\'enyi random graphs $G(n, p)$, which is obtained from a complete graph of $n$ vertices by keeping each edge with probability $p$ or deleting it, and each edge is independent of all other edges. 
	For the largest connected component of $G(n, 1/n)$, the mixing time and the hitting time are both $\Theta(n)$ \cite{Nachmias2008CriticalRG}, which means that our algorithm has a quadratic speedup compared with the classical case and behaves well in large sparse graph such as the real-world networks. The same quadratic speedup will be obtained in some common graphs and we list some examples as follows. 
	For the sake of brevity, the part about $\varepsilon$ in each item is omitted, and all have a logarithmic relationship with $\varepsilon$.
	\begin{table}[h!]
		\begin{center}
			\resizebox{\linewidth}{!}{
				\begin{tabular}{c c c c}
					\hline
					\textbf{Graph} & \textbf{Classical mixing time} & \textbf{Our result} & \textbf{Best previous result}\\
					\hline	
					balanced r-tree &  $n$  &  $\sqrt{n\log n}$  &  $\sqrt{n}\log^{3/2} n$ \\
					barbell &  $n^3$  &  $n^{3/2}$  &  $n^{3/2}\log n$ \\
					circle &  $n^2$  &  $n$  &  $n$    \\
					necklace &  $n^2$  &  $n$  &  $n\log n$  \\ 
					two cliques glued at a single vertex &  $n^2\log n$  &  $n$  &  $n\log n$  \\ 
					G(n, 1/n) &  $n$  &  $\sqrt{n}$  &  $\sqrt{n}\log n$  \\ 
					\hline
			\end{tabular}}
			\caption{The tight bound quantitative results of many graphs.  All above results are $\Theta(\cdot)$, where the detailed definition of the first four graphs and the corresponding hitting time and mixing time results can be found in \cite{Reversible}, the fifth graphs can be found in \cite{Apers2021AUF} and the last graphs is in \cite{Nachmias2008CriticalRG}. The corresponding  best previous quantum  results are from the result of \cite{Krovi2015QuantumWC, PhysRevA.102.022423}.}
		\end{center}
	\end{table}

	Certainly, it is not yet possible to achieve quadratic acceleration for any reversible Markov chain now and we can consider the problem in the future.
	\section{Acknowledgements}
	We thank the support of National Natural Science Foundation of China (Grant No.61872352), and Program for Creative Research Group of National Natural Science Foundation of China (Grant No. 61621003).

\end{document}